\documentclass[aps,prl,epsfig,floats,twocolumn,superscriptaddress,amssymb,amsmath,showpacs]{revtex4}
\usepackage{graphicx}
\begin{document}
\title{Elastic Correlations in Nucleosomal DNA Structure}

\author{Farshid Mohammad-Rafiee}

\affiliation{Isaac Newton Institute for Mathematical Sciences,
Cambridge, CB3 0EH, UK}

\affiliation{Institute for Advanced Studies in Basic Sciences,
Zanjan 45195-159, Iran}

\author{Ramin Golestanian}

\affiliation{Isaac Newton Institute for Mathematical Sciences,
Cambridge, CB3 0EH, UK}

\affiliation{Institute for Advanced Studies in Basic Sciences,
Zanjan 45195-159, Iran}

\date{\today}

\begin{abstract}

The structure of DNA in the nucleosome core particle is studied
using an elastic model that incorporates anisotropy in the bending
energetics and twist-bend coupling. Using the experimentally
determined structure of nucleosomal DNA [T.J. Richmond and C.A.
Davey, Nature {\bf 423}, 145 (2003)], it is shown that elastic
correlations exist between twist, roll, tilt, and stretching of
DNA, as well as the distance between phosphate groups. The
twist-bend coupling term is shown to be able to capture these
correlations to a large extent, and a fit to the experimental data
yields a new estimate of $G=25$ nm for the value of the twist-bend
coupling constant.

\end{abstract}

\pacs{87.15.-v, 87.15.La, 87.14.Gg, 82.39.Pj}

\maketitle

The DNA in eukaryotes is tightly bound to an equal mass of histone
proteins, forming a repeating array of DNA-protein complexes
called nucleosomes \cite{Cell}. A stretch of 147 base pair (bp)
DNA is wrapped in 1.84 left-handed superhelical turns around the
histone octamer that forms the nucleosome core particle, which is
connected via a linker DNA to the next core particle. Each
nucleosome core is a tiny sized spool with a radius of 5 nm and a
height of 6 nm \cite{Luger}. The wrapped DNA-histone octamer
complex is essentially ubiquitous in nature and has a major role
in many cell life processes such as gene expression and
transcription \cite{Seacker}.

In a recent high precision measurement, Richmond and Davey have
determined the structure of the 147 base pair DNA in the
nucleosome core particle with $1.9 \; {\rm \AA}$ resolution
\cite{Richmond}. They have observed that the structure of the bent
DNA segment is modulated in the curvature, roll, and tilt, and
that the twist structure appears to be most affected by the
specific interactions with the protein substrate. This experiment
provides a wealth of information about the conformational
structure of such highly bent and strongly interacting DNA, among
which we can highlight a number of quantitative observations: (1)
the period of modulation in curvature is set by half of the DNA
pitch $\sim 5$ bp, where either the major or the minor grooves
face the histone octamer, (2) roll appears to have the main
contribution to the curvature, as it is favored over tilt by
$1.9:1$, and (3) the DNA segment is stretched by about 1-2 bp as
compared to its unbent conformation \cite{Richmond}.

\begin{figure}
\includegraphics[width=.75\columnwidth]{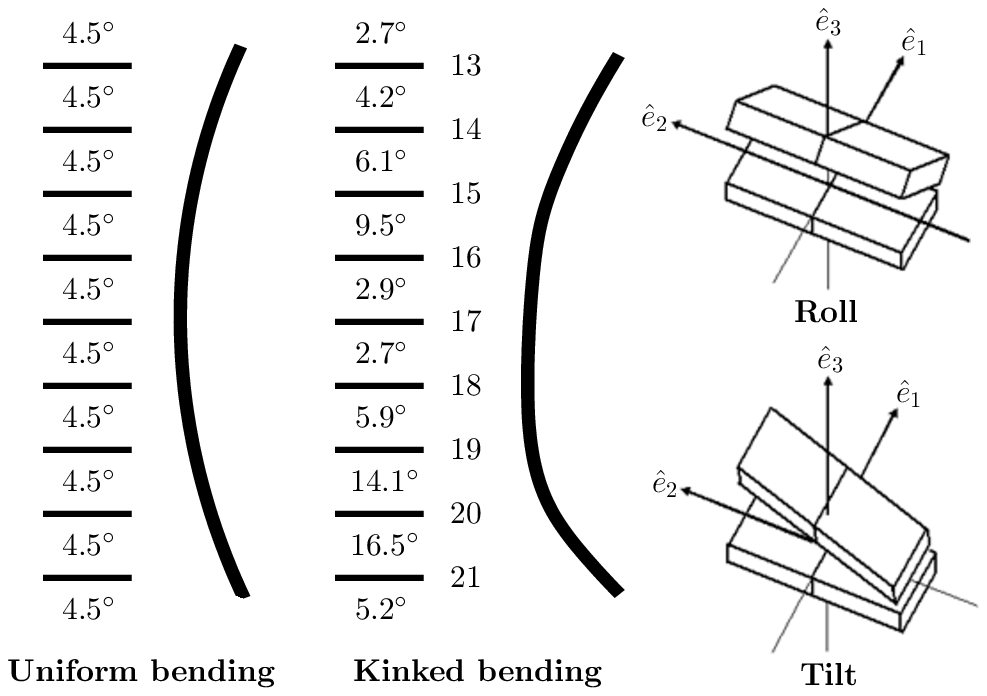}
\caption{The conformation of a segment of the nucleosomal DNA,
between the 13th and the 21st base pairs, is compared with that of
a simple homogenous elastic rod that bends uniformly. The bending
angles between two consecutive base pairs give a measure of the
non-uniformity in the shape of the highly bent DNA segment. Note
the sharp bends (kinks) near the 15th and the 20th base pairs. A
schematic definition of the bending components Roll and Tilt is
presented in the right panel.} \label{fig:conformation}
\end{figure}

The conformational properties of relatively long DNA segments, as
well as their elastic response to mechanical stresses such as
pulling forces and torques, have been successfully studied using a
coarse-grained elastic description
\cite{Benham,Tanaka,Marko,Kamien1,Fain1} that could take into
account thermal fluctuations \cite{Marko-PRE,Bouchiat1,Panyukov1}.
Other approaches for studying DNA structure include
first-principle computer simulations \cite{Schlick,Olson2} and
phenomenological modelings using base-stacking interactions
\cite{Kamien2,Mergel}. In light of the recent experimental
determination of the nucleosomal DNA structure, a corresponding
theoretical analysis is called for, and one naturally wonders
which of the above approaches could more easily accommodate the
additional complications due to the high degree of bending and the
specific DNA-protein interactions.

Here, we attempt to use an augmented elastic description to
account for a number of observations made by Richmond and Davey.
We consider an elastic energy expression that includes anisotropic
bending rigidities and twist-bend coupling. We show that the
anisotropic bending elasticity is responsible for the modulations
in curvature with the period of 5 bp \cite{Farshid1}, and
calculate the shape of DNA, an example of which is shown in Fig.
\ref{fig:conformation}. Using an analysis of the experimental data
of Ref. \cite{Richmond}, we show that the specific features in
twist and bend are correlated to a large extent through the
twist-bend coupling, and extract an estimate of $G=25$ nm for the
twist-bend coupling constant that best describes this correlation.
We calculate the components of curvature as well as the axial
strain using the experimental values for twist as input, and find
that roll is favored over tilt by $1.7:1$, and that the overall
stretching of DNA is about 1 bp, both in encouraging quantitative
agreement with the experiment. We have also studied another
parameter called $\Delta {\rm PP_a}$, which shows the
super-helix-component of the difference of the lengths of the two
phosphodiester chains \cite{Richmond}, and found good agreement
with the experimental data.

\begin{figure}
\begin{center}
\includegraphics[width=.68\columnwidth]{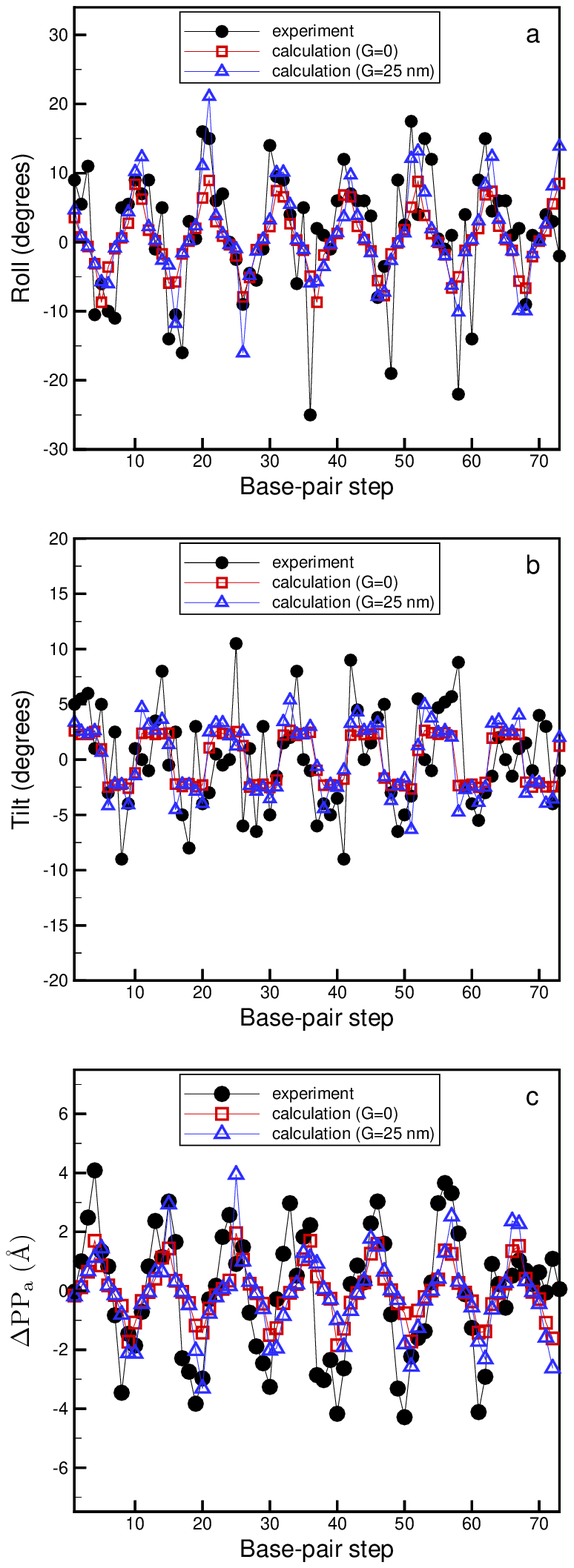}
\end{center}
\caption{(a) Roll, (b) tilt, and (c) the super-helix-component of
the difference of the lengths of the two phosphodiester chains.
The filled circles are the experimental data taken from Ref.
\cite{Richmond}, the hollow triangles (squares) are calculated
points with (without) the twist-bend coupling.}
\label{fig:roll-tilt}
\end{figure}

To study the structure of the nucleosomal DNA, we consider a
simple model in which the molecule is represented as an elastic
rod \cite{Farshid1}. The rod is parameterized by the arclength $s$
and at each point, an orthonormal basis is defined with the unit
vectors $\hat{e}_1(s)$, $\hat{e}_2(s)$, and $\hat{e}_3(s)$, where
$\hat{e}_1$ shows the direction from minor groove to major groove,
and $\hat{e}_3(s)$ is the unit tangent to the axis (see Fig.
\ref{fig:conformation}). Note that due to the helical structure of
DNA, $\hat{e}_1(s)$ and $\hat{e}_2(s)$ rotate with the helix. The
deformation of the double helix is characterized by the angular
strains $\Omega_{1,2}(s)$ corresponding to bending in the plain
perpendicular to $\hat{e}_{1,2}(s)$ and $\Omega_3(s)$
corresponding to twist and torsion. We should mention that each
slice of the rod is labeled by $s$, which is corresponding to arc
length along the unstressed helix axis and so always changes from
$0$ to $L$. The actual arc length along the deformed axis is given
by $ds^\prime$, which in terms of axial strain $\alpha(s)$ one
have $ds^\prime = [1 + \alpha(s)] ds$ \cite{note1}. The elastic
energy for the deformation of DNA in units of thermal energy is
written as \cite{Marko,Kamien1}
\begin{eqnarray}
\frac{E_{\rm DNA}}{k_{\rm B} T}&=&\frac{1}{2} \int_0^L d s \left[
A_1
\Omega_1^2 + A_2 \Omega_2^2 + C(\Omega_3 - \omega_0)^2 \right. \nonumber \\
&&+ \left.B \omega_0^2 \alpha^2 + 2 D \omega_0 (\Omega_3 -
\omega_0) \alpha \right. \nonumber \\
&&+ \left.2 \widetilde{G} \omega_0 (\Omega_3 - \omega_0) \Omega_2
\right], \label{E1}
\end{eqnarray}
where $A_1$ and $A_2$ are the bending rigidities for the ``hard''
and ``easy'' axes of DNA cross section, $C$ is the twist rigidity,
$B$ is the stretch modulus, $D$ is the twist-stretch coupling, and
$G \equiv \widetilde{G} \omega_0$ is the twist-bend coupling. The
spontaneous twist of the helix is defined via its pitch $P$ as
$\omega_0 = 2 \pi/P$, and for B-DNA we have $\omega_0 = 1.85 \;
{\rm nm}^{-1}$. Note that the twist-bend coupling $G$ can be ruled
out by symmetry ($\psi$ to $-\psi$) for a non-chiral rod
$\omega_0=0$, and thus its presence is a direct consequence of a
spontaneous twist structure, as manifested by the form $G \equiv
\widetilde{G} \omega_0$.

The local strains $\Omega_i$ can be written in terms of the
curvature $\kappa(s)$, the torsion $\tau(s)$, and the twist angle
$\psi(s)$ as $\Omega_1 = \kappa \sin \psi$, $\Omega_2 = \kappa
\cos \psi$, and $\Omega_3 = \tau + \partial_s {\psi}$
\cite{Marko}. It can be shown that the mean curvature and torsion
imposed by the nucleosomal structure are $\kappa_{\rm av} = R /
\left( R^2 + \nu^2\right)$ and $\tau_{\rm av} = - \nu / \left( R^2
+ \nu^2\right)$, where $R$ is the radius of the nucleosome and $2
\pi \nu$ denotes the pitch of the wrapped DNA around the histone
octamer. For the nucleosome case, $R \simeq 41.9 \; \AA$ and $2
\pi \nu \simeq 25.9 \; \AA$ \cite{Richmond}, which yields
$\tau_{\rm av}/\kappa_{\rm av}=0.098$. Since typical values of
$\partial_s {\psi}$ is of the order of $\omega_0$, we can estimate
the relevance of torsion with respect to twist by the ratio
$\tau_{\rm av}/\omega_0=0.012$. These estimates justify neglecting
torsion with respect to twist and curvature. After defining $A
\equiv \frac{1}{2} (A_1 + A_2)$, and $A^\prime \equiv \frac{1}{2}
(A_1-A_2)$, we can write Eq. (\ref{E1}) as
\begin{eqnarray}
\frac{E_{\rm DNA}}{k_{\rm B} T}&=&\frac{1}{2} \int_0^L d s
\left[(A - A^\prime \cos 2 \psi) \kappa^2
+ B \omega_0^2 \alpha^2 \right. \nonumber \\
&&+\left. C(\partial_s \psi- \omega_0)^2 + 2 D \omega_0
(\partial_s \psi-\omega_0) \alpha \right. \nonumber \\
&&+\left. 2 G (\partial_s{\psi} - \omega_0) \kappa \cos
\psi\right]. \label{E}
\end{eqnarray}
For the coupling parameters, we use $A=50$ nm \cite{Marko-PRE},
$A^\prime=30$ nm \cite{Olson}, $B=78$ nm \cite{Wang}, $C=75$ nm
\cite{Vologodskii}, and $D=15$ nm \cite{Marko3}. Since the only
estimate available for the value of the twist-bend coupling
constant $G$ has been rather indirect \cite{Marko}, we treat $G$
as a tuning parameter and find its value by fitting to the
experimental data of Ref. \cite{Richmond}.

To find the shape of the DNA, we need to incorporate the
interactions with the binding substrate. The wrapping of DNA by an
overall angle of $\Theta=2 \pi \times 1.84$ around the protein
octamer can be imposed as a global constraint, which reads
$\int_0^L ds (1 + \alpha) \kappa = \Theta$.
To take account of the specific and local interactions, we assume
that it is mostly the twist degree of freedom that is affected by
those interactions \cite{Schiessel-Review}; a view that is
supported by the experimental observations of Ref.
\cite{Richmond}. This allows us to write the interaction energy
term, which is the final addition to the energy expression for the
DNA conformation Eq. (\ref{E}), as $V_{\rm DNA-Histone}[\psi]$. We
can now minimize the sum of $E_{\rm DNA}+V_{\rm
DNA-Histone}[\psi]$ with respect to $\kappa$, $\alpha$, and
$\psi$, subject to the wrapping constraint
above.  This yields expressions for the
curvature and the axial strain as functions of the twist angle as
\begin{eqnarray}
\kappa &=& \frac{(\partial_s{\psi} - \omega_0) \left[\mu \omega_0
D + B G \omega_0^2 \cos \psi \right] - \mu B \omega_0^2}{\mu^2 - B
\omega_0^2 \left( A - A^\prime \cos
2 \psi \right) }, \label{kappa} \\
\alpha &=& \frac{(\partial_s{\psi} - \omega_0) \left[ D \omega_0
\left( A- A^\prime \cos \psi\right) + \mu G \cos \psi \right]-
\mu^2}{\mu^2 - B \omega_0^2 \left( A - A^\prime \cos
2 \psi \right) }, \nonumber \\
\label{alpha}
\end{eqnarray}
where $\mu$ is the Lagrange multiplier for the constraint, as well
as an equation for the twist that contains unknown interaction
terms arising from $V_{\rm DNA-Histone}[\psi]$. Instead of
elaborating on the possible forms of this interaction, we simplify
the procedure by directly reading off the twist angle from the
experimental results of Richmond and Davey. This allows us to make
a direct comparison between the calculated values for the
curvature and stretching and the corresponding experimentally
measured values for them, and hence put the elastic model to a
stringent test.

\begin{figure}
\includegraphics[width=.7\columnwidth]{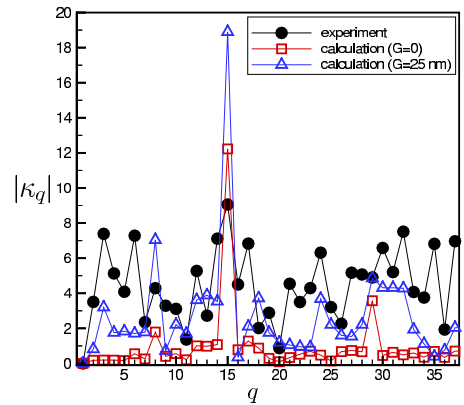}
\caption{The Fourier transform $\kappa_q$ of the curvature. The
filled circles are the experimental data taken from Ref.
\cite{Richmond}, the hollow triangles (squares) are calculated
points with (without) the twist-bend coupling. The peak at
$q=15.6$ corresponds to a periodicity of 5 bp. (A large peak at
zero corresponding to the average of the curvature is eliminated
to enhance the resolution of the figure.)} \label{fig:kappa}
\end{figure}

The twist angle is obtained from the twist strain by integration,
and the corresponding integration constant is chosen by noting
that at the dyad axes of the nucleosome the major groove-minor
groove direction is perpendicular to the surface of the nucleosome
core particle. This means that the first base pair in either the
left or the right half of the nucleosomal DNA should have an
offset twist angle of $\psi_0 = 2 \pi \times \frac{1 \;{\rm
bp}}{10 \;{\rm bp}}=36^\circ$. Finally, we note that following
Ref. \cite{Richmond} the calculations are performed for half of
the DNA length corresponding to 73 base pairs, to make a direct
comparison possible.

In Figs. \ref{fig:roll-tilt}a and \ref{fig:roll-tilt}b, we have
plotted the calculated roll ${\cal R}= \frac{180}{\pi} b \Omega_2
=\frac{180}{\pi} \kappa b \cos \psi$ and tilt ${\cal
T}=\frac{180}{\pi} b \Omega_1 =\frac{180}{\pi} \kappa b \sin \psi$
\cite{Dickerson} as functions of the position of the base pairs,
together with the experimental data of Ref. \cite{Richmond}, where
$b=3.4 \; {\rm \AA}$ is the base-pair step for B-DNA. The two
quantities appear to be modulated with a near periodicity of 10
bp, which is imposed by the near periodicity in $\psi$. The best
fit to the experimental data yields $G=25$, which results in the
roll being favored over tilt by $1.7:1$, to be compared with the
experimental value of $1.9:1$. This shows that in such a highly
bent structure, the DNA prefers by a ratio of nearly 2 to 1 to use
the bending over the easy axis as opposed to the hard one.

To make a more refined quantitative comparison with the
experiment, we consider the net curvature (as opposed to its
components) and take its Fourier transform defined as
$\kappa_q=\frac{1}{\sqrt{n}} \sum_{s=1}^n \, \kappa_s e^{2 \pi i
(s-1) (q-1) / n}$ for a list $\kappa_s$ of length $n$ ($n=73$
here), to better resolve its features. A plot of the absolute
value of the curvature Fourier transform is shown in Fig.
\ref{fig:kappa}, comparing the experimental data with the
calculated ones. Note that the absolute value of the Fourier
transform is symmetric with respect to the transformation $q \to
n-q$, hence only the first half of the plot is shown. The Fourier
transform of the curvature shows a distinct peak at
$q=\frac{73}{5}+1=15.6$ corresponding to a periodicity of 5 bp,
which is a result of the anisotropic bending elasticity
\cite{Farshid1}. Moreover, Fig. \ref{fig:kappa} shows that while
the simple elastic description fails to account for the detailed
features of the curvature without a twist-bend coupling, once
equipped with such a term it can give a considerably improved
account, with the best fit corresponding to $G=25$ nm. We also
note that using the values for the curvature the shape of the bent
DNA can be determined upon integration. We have provided such an
example in Fig. \ref{fig:conformation}, where the presence of two
kinks at the distance of 5 base pairs from each other can be
visibly noted.

\begin{figure}
\includegraphics[width=.58\columnwidth]{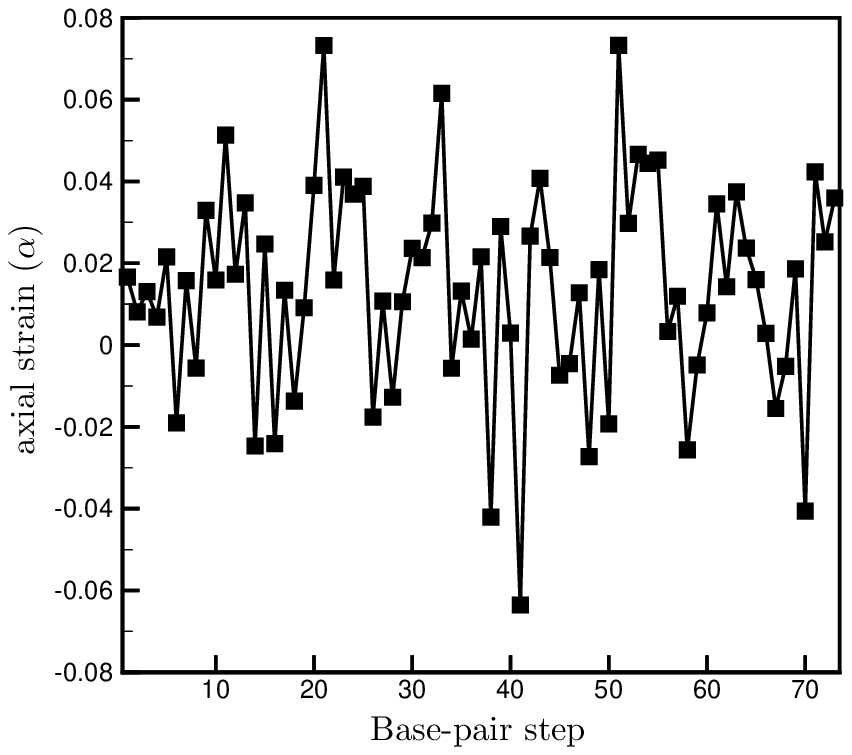}
\caption{The calculated axial strain as a function base-pair
position. While detailed experimental data is lacking for the
stretching at each base pair, the overall stretching obtained by
integration over the above data gives 1 bp in agreement with the
observations of Ref. \cite{Richmond}.} \label{fig:alpha}
\end{figure}

Using Eq. (\ref{alpha}), one can also determine the stretching of
the nucleosomal DNA. In Fig. \ref{fig:alpha}, the axial strain
$\alpha$ is plotted as a function of the base-pair position. A
positive (negative) value of $\alpha$ shows a stretching
(compression) for the corresponding base-pair. The overall length
of the DNA in the nucleosome can be found as $\int_0^L \; ds \;
(1+\alpha) = 148$ bp, which suggests that the DNA in the
nucleosome is stretched by about 1 bp, in agreement with the
observations of Richmond and Davey \cite{Richmond}.

We have also calculated the difference between the components of
the phosphate-phosphate distances lying parallel to the path of
the superhelix $\Delta {\rm PP_a} =(\vec{\ell}_1 - \vec{\ell}_2)
\cdot \hat{e}_3(s)$, where $\vec{\ell}_i$ gives the
phosphate-phosphate distances on the $i$th strand. Using the
geometrical definitions, we find
$\Delta {\rm PP_a} = \frac{2 d \kappa}{\Omega^2}\;  \cos\psi \;
(\partial_s \psi)\; (1 - \cos b \Omega)$,
where $\Omega \equiv \sqrt{\kappa^2 + (\partial_s \psi)^2}$, and
$d=2$ nm is the diameter of the undeformed DNA. The calculated
values of $\Delta {\rm PP_a}$ are shown in Fig.
\ref{fig:roll-tilt}c, which are in good agreement with the
experimental values taken from Ref. \cite{Richmond}.

We have also examined the effect of other elastic couplings---such
as the bend-stretch coupling---by trying to fit to the
experimental data, and have found no significant effect. Higher
order elastic terms such as the cubic terms in curvature etc.
\cite{Marko}, are expected to add corrections of the order of
$\kappa d$, which become (marginally) important in the highly bent
(kink) regions. While the addition of such terms would definitely
help improve the results quantitatively, the fact that it will
introduce more unknown coupling constants make such a direction
not particularly appealing. Moreover, there are also other
structural properties of the nucleosomal DNA such as shift and
slide \cite{Dickerson}, which appear to be beyond such a simple
elastic description. This suggests that a more promising direction
for an improved theory that can better describe the conformation
of nucleosomal DNA is a generalization of the base-stacking model
of O'Hern et al. \cite{Kamien2,ToBePub}.

In conclusion, we have shown that an elastic theory that takes
into account anisotropic bending and twist-bend coupling can
account to a considerable degree for the observed structure of the
nucleosomal DNA. Since a full microscopic computer simulation of
such a large DNA-protein complex appears to be out of reach with
the computational power at hand, such simplified phenomenological
approaches could be helpful in understanding the structural
properties of biomolecules.


We thank R. Bruinsma, H. Flyvbjerg, T.B. Liverpool, P. D. Olmsted,
Z.-C. Ou-Yang, W.C.K. Poon, M. Rao, H. Schiessel, and A. Travers
for very helpful discussions.



\begin{thebibliography}{99}
\bibitem{note1}
The fact that the backbone is considered as inextensible does not
contradict with the introduction of the axial strain, which as a
field defined on the inextensible coordinate system of the
backbone describes the extension of the rod.
\end{thebibliography}
\end{document}